\documentclass[camera]{jpaper}

\usepackage[nocompress]{cite}
\usepackage{algorithmic}
\usepackage{array}

\makeatletter
\let\MYcaption\@makecaption
\makeatother

\usepackage[font=footnotesize]{subcaption}

\makeatletter
\let\@makecaption\MYcaption
\makeatother

\usepackage{fixltx2e}
\usepackage{dblfloatfix}
\usepackage[nolessnomore, italic]{mathastext}
\usepackage[T1]{fontenc}
\usepackage[usenames,dvipsnames,svgnames,table]{xcolor}
\usepackage[normalem]{ulem}
\usepackage{enumitem}
\usepackage{setspace}
\usepackage{indentfirst}
\usepackage{footmisc}
\usepackage{fancyhdr}
\usepackage{authblk}
\usepackage[us,12hr]{datetime}
\usepackage[keeplastbox]{flushend}
\usepackage[hidelinks]{hyperref}
\usepackage{xspace}

\newboolean{publicversion}
\setboolean{publicversion}{true}

\ifthenelse{\boolean{publicversion}}{
  \pagenumbering{arabic}
  \newcommand{\grumbler}[2]{}
  \newcommand{\assign}[1]{}
  \newcommand{\respond}[3]{}
  \newcommand{\changesI}[0]{}
  \newcommand{\changesII}[0]{}
  \newcommand{\changesIII}[0]{}

}{
  \pagenumbering{arabic}
  \newcommand{\grumbler}[2]{\textcolor{blue}{\bf #1: #2}}
  \newcommand{\assign}[1]{\textcolor{purple}{\bf RESPONSIBLE: #1}}
  \newcommand{\respond}[3]{\textcolor{#1}{\bf #2-response: #3}}
  \newcommand{\changesI}[0]{}
  \newcommand{\changesII}[0]{}
  \newcommand{\changesIII}[1]{\textcolor{BrickRed}{#1}}

}

\newcommand{\titleShortSMS}[0]{SMS\xspace}
\newcommand{\titleLongSMS}[0]{Staged Memory Scheduler\xspace}
\newcommand{\titleShort}[0]{SMS\xspace}
\newcommand{\paragraphbe}[1]{{\textbf{#1}}}

\newcommand{\batch}[0]{batch\xspace}
\newcommand{\Batch}[0]{Batch\xspace}
\newcommand{\batches}[0]{batches\xspace}

\newcommand{\ignore}[1]{}

\newif\ifcameraready
\camerareadytrue

\newcommand{\versionnum}[0]{5.1}

\ifcameraready
\else
\fi

\begin{document}
%
\title{High-Performance and Energy-Efficient Memory Scheduler Design\\for Heterogeneous Systems}



\newcommand{\authspace}[0]{\qquad}
\newcommand{\affilspace}[0]{\qquad}

\author{
    Rachata Ausavarungnirun$^1$ \authspace
    Gabriel H. Loh$^2$
    \vspace{2pt}\\
    Lavanya Subramanian$^{3,1}$ \authspace
    Kevin Chang$^{4,1}$ \authspace
    Onur Mutlu$^{5,1}$}
\affil{\it
	$^1$Carnegie Mellon University\affilspace
	$^2$AMD Research\affilspace
	$^3$Intel Labs\affilspace
	$^4$Facebook \affilspace
    $^5$ETH Z{\"u}rich}

\maketitle


\begin{abstract}

This paper summarizes the idea of the Staged Memory Scheduler (SMS), which was
published at ISCA 2012~\cite{sms}, and examines the work's
significance and future potential.
When multiple processor cores (CPUs) and a GPU integrated together on the same
chip share the off-chip DRAM, requests from the GPU can heavily interfere with
requests from the CPUs, leading to low system performance and starvation of
cores. Unfortunately, state-of-the-art memory scheduling algorithms are
ineffective at solving this problem due to the very large amount of \changesI{GPU memory traffic},
unless a very large and costly request buffer is employed to provide these
algorithms with enough visibility across the global request stream.

\changesI{Previously-proposed} memory controller (MC) designs use a single monolithic
structure to perform three main tasks. First, the MC attempts to schedule
together requests to the same DRAM row to increase row buffer hit rates.
Second, the MC arbitrates among the requesters (CPUs and GPU) to optimize for
overall system throughput, average response time, fairness and quality of
service. \changesI{Third}, the MC manages the low-level DRAM command \changesII{scheduling} to \changesII{complete requests while ensuring} 
compliance with all DRAM timing and power constraints.

This paper proposes a fundamentally new approach, \changesI{called the Staged Memory Scheduler (SMS), which} decouples the three
primary MC tasks into three significantly simpler structures that together
improve system performance and fairness. Our three-stage MC first groups
requests based on row buffer locality. This grouping allows the second stage to
focus only on inter-application scheduling decisions. These two stages enforce
high-level policies regarding performance and fairness, and therefore the last
stage can \changesI{use} simple per-bank FIFO queues (\changesI{i.e., there is no need for} further command
reordering within each bank) and straightforward logic that deals only with
the low-level DRAM commands and timing.

We evaluated the design trade-offs involved 
and compared it against four state-of-the-art MC designs. Our evaluation shows that
\titleShortSMS  provides  41.2\% performance improvement and 4.8$\times$ \changesII{fairness improvement} compared to the best previous state-of-the-art technique, while
\changesII{enabling} a design that is significantly less complex and more power-efficient to implement.

Our analysis and proposed scheduler have inspired significant research on \changesIII{(1) predictable and/or
deadline-aware memory
scheduling~\cite{kim-rtas2014, kim2016, usui-squash, usui-dash,
jishen-firm,bliss,bliss-tpds,mise} and (2) memory scheduling for heterogeneous systems~\cite{usui-squash,usui-dash,wang-pact14,pattnaik-pact2016}}.

\ignore{
In addition, only
the last stage of \titleShortSMS needs to deal with correctness and power issues,
the design of \titleShortSMS is significantly less complex to implement.
}
\end{abstract}

\section{Introduction}
\label{prelim-sms}

As the number of cores continues to increase in modern chip multiprocessor (CMP)
systems, the DRAM memory system \changesII{has become} a critical shared resource~\cite{superfri,imw2013}.
Memory requests from multiple cores interfere with each other, and this
inter-application interference is a significant impediment to individual
application and overall system performance. \changesII{Various works} on application-aware
memory scheduling \cite{atlas,tcm,stfm, parbs} \changesII{address} the problem by
making the memory controller aware of application characteristics and
appropriately prioritizing memory requests to improve system performance and
fairness.

Recent heterogeneous CPU-GPU systems~\cite{bobcat,sandybridge,amd-fusion,apu,kaveri,haswell,amdzen,skylake,powervr,arm-mali,tegra,tegrax1} present an additional
challenge by introducing integrated graphics processing units (GPUs) on
the same die with CPU cores. GPU applications typically demand significantly
more memory bandwidth than CPU applications due to the GPU's capability of
executing a large number of \changesII{concurrent} threads~\cite{solomon62,senzig-afips65,crane-ec65,hellerman-ec66,cdc7600,cdcstar,illiac,cray1,cdc6600,cdc6600-2,hep,masa-fmt,april-fmt,tera-mta,amd-fusion,apu,kaveri,haswell,amdzen,skylake,powervr,arm-mali,tegra,tegrax1,fermi,kepler,maxwell,pascal,amdr9,radeon,vivante-gpgpu}. GPUs use
\emph{single-instruction multiple-data} (SIMD) pipelines to concurrently execute multiple threads~\cite{flynn}.
\changesII{In a GPU, a group of threads executing} the same instruction is called a
\emph{wavefront} or \emph{warp}, \changesII{and threads in a warp are} executed in lockstep. When a wavefront stalls on a memory instruction, the GPU core
hides this memory access latency by switching to another wavefront to avoid stalling
the pipeline. Therefore, there can be thousands of outstanding memory requests
from across all of the wavefronts. This is fundamentally more memory intensive than CPU
memory traffic, where each CPU application has a much
smaller number of outstanding requests due to the sequential execution model
of CPUs.

\begin{figure*}[h]
\centering
\includegraphics[width=1.7\columnwidth]{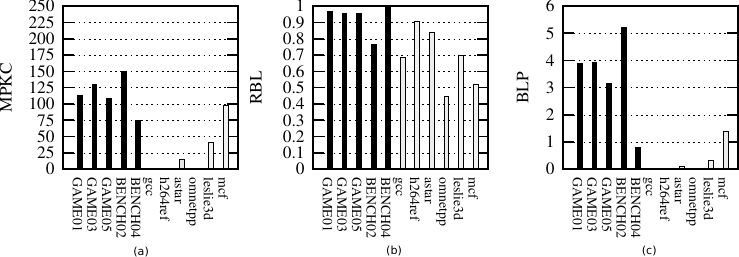}
\caption{GPU memory characteristics. (a) Memory intensity, measured by memory requests
per thousand cycles, (b) row buffer locality, measured by the fraction
of accesses that hit in the row buffer, and (c) bank-level parallelism. Reproduced from~\cite{sms}.}
\label{fig:intensity}
\end{figure*}

Figure~\ref{fig:intensity} (a) shows the memory request rates for a
representative subset of our GPU applications and the most memory-intensive
SPEC2006 (CPU) applications, as measured by memory requests per thousand cycles
when each application runs alone on the system. The raw bandwidth demands \changesII{(i.e., memory request rates)} of
the GPU applications are often multiple times higher than the SPEC benchmarks.
Figure~\ref{fig:intensity} (b) shows the row buffer hit rates (also called
\emph{row buffer locality} or RBL~\cite{memattack}). The GPU applications show consistently high
levels of RBL, whereas the SPEC benchmarks exhibit more variability. The GPU
programs have high levels of spatial locality, often due to access patterns
related to large sequential memory accesses (e.g., frame buffer updates).
Figure~\ref{fig:intensity}(c) shows the \emph{bank-level parallelism (BLP)~\cite{parbs,cjlee-micro09}, which is the
average number of parallel memory requests that can be issued to different DRAM banks,} for
each application, with the GPU programs consistently making use of \emph{four} banks at
the same time.

In addition to the high-intensity memory traffic of GPU applications, there are
other properties that distinguish GPU applications from CPU applications. \changesI{Prior work}~\cite{tcm} 
observed that CPU applications with streaming access
patterns typically exhibit high RBL but low BLP, while applications with less
uniform access patterns typically have low RBL but high BLP.

In contrast, GPU applications have {\em both} high RBL and
high BLP. The combination of high memory intensity, high RBL and high BLP means
that the GPU will cause significant interference to other applications
across all banks, especially when using a memory scheduling algorithm
that preferentially favors requests that result in row buffer hits \changesII{(e.g.,~\cite{frfcfs-patent,fr-fcfs})}.

\begin{figure}
\centering
\includegraphics[width=\columnwidth]{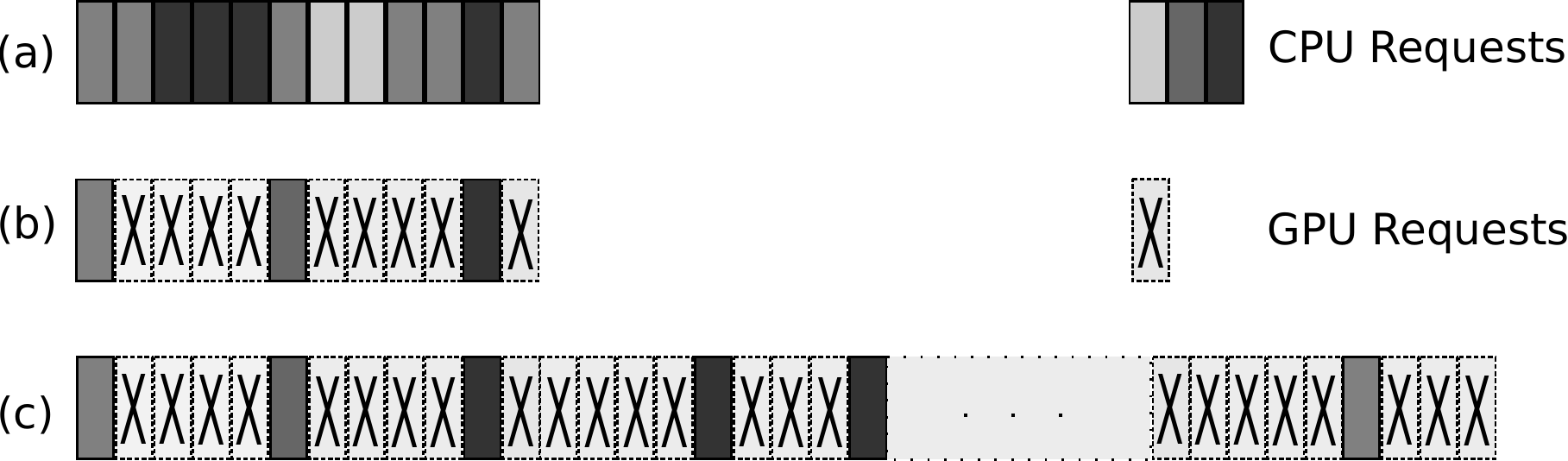}
\caption{\changesII{Example of the limited visibility of the memory controller}. (a) CPU-only information, (b) Memory controller's visibility, (c) Improved visibility. Adapted from~\cite{sms}.}
\label{fig:visibility}
\end{figure}

Recent memory scheduling research has focused on
memory interference between applications in CPU-only scenarios. These
past proposals are built around a single centralized request buffer at each memory
controller (MC). The scheduling algorithm implemented in the memory controller
analyzes the stream of requests in the centralized request buffer to
determine application memory characteristics, decides on a priority for each
core, and then enforces these
priorities. Observable memory characteristics may include the number of requests
that result in row buffer hits, the bank-level parallelism of each core, memory request
rates, overall fairness metrics, and other information.
Figure~\ref{fig:visibility}(a) shows the CPU-only scenario where the request
buffer only holds requests from the CPUs. In this case, the memory controller sees a number of
requests from the CPUs and has visibility into their memory behavior. On the
other hand, when the request buffer is shared between the CPUs and the GPU, as
shown in Figure~\ref{fig:visibility}(b), the large volume of requests from the
GPU occupies a significant fraction of the memory controller's request buffer, thereby limiting the
memory controller's visibility of the CPU applications' memory \changesII{characteristics}.

One approach to increasing the memory controller's visibility across a larger
window of memory requests is to increase the size of its request buffer. This
allows the memory controller to observe more requests from the CPUs to better
characterize their memory behavior, as shown in
Figure~\ref{fig:visibility}(c). For instance, with a large request buffer, the
memory controller can identify and service multiple requests from one CPU core to the same row such that
they become row buffer hits, however, with a small request buffer as shown in
Figure~\ref{fig:visibility}(b), the memory controller may not even see these requests at the same time
because the GPU's requests have occupied the majority of the entries.

Unfortunately, very large request buffers impose
significant implementation challenges, including the die area for the larger
structures and the additional circuit complexity for analyzing so many requests,
along with the logic needed for assignment and enforcement of \changesII{priorities~\cite{bliss,bliss-tpds}}. Therefore, while building
a very large, centralized memory controller request buffer could \changesII{perhaps} lead to \changesII{reasonable} memory
scheduling decisions, the approach is unattractive due to the resulting area,
power, timing and complexity costs.

In this work, we propose the \titleLongSMS (\titleShortSMS), a decentralized
architecture for memory scheduling in the context of integrated multi-core CPU-GPU systems. The key idea in \titleShortSMS is to decouple the various functional \changesII{tasks}
of memory controllers and partition these tasks \emph{across} several simpler hardware
structures which operate in a staged fashion.
The three primary functions of the memory controller, which map to the
three stages of our proposed memory controller architecture, are:

\begin{enumerate}
\item Detection of basic \changesI{intra-application} memory characteristics (e.g.,
row buffer locality).
\item Prioritization across applications (CPUs and GPU) and enforcement
of policies to reflect the priorities.
\item Low-level command scheduling (e.g., activate, precharge, read/write), enforcement
of \changesII{DRAM} device timing constraints (e.g., t$_\text{RAS}$, t$_\text{FAW}$, etc.), and
\changesII{resolution of} resource conflicts (e.g., data bus arbitration).\footnote{\changesI{We refer the reader to our prior works~\cite{atlas,tcm,salp,tl-dram,al-dram,chargecache,raidr,dsarp,lisa,seshadri2013rowclone,chang-sigmetric17,ava-dram,ambit,liu-isca2013, chang-sigmetric16, softmc, kim-cal2015, lee-pact2015, donghyuk-stack, kim-isca2014, patel-isca2017, kim-hpca2018} for a detailed background on DRAM.}}
\end{enumerate}

Our specific \titleShortSMS implementation makes widespread use of distributed FIFO structures
to maintain a very simple implementation, but at the same time \titleShortSMS can provide
fast service to low memory-intensity (likely latency-sensitive) applications and effectively
exploit row buffer locality and bank-level parallelism for high memory-intensity (bandwidth-demanding) 
applications. While \titleShortSMS provides a specific implementation, our
staged approach for memory controller organization provides a general framework for exploring scalable
memory scheduling algorithms capable of handling the diverse memory needs of
integrated \changesII{heterogeneous processing} systems of the future \changesII{(e.g., systems-on-chip
that contain CPUs, GPUs, and accelerators)}.

\section{\titleLongSMS Design}

{\bf Overview:} Our proposed \titleLongSMS~\cite{sms} architecture
introduces a new memory controller (MC) design that provides 1) scalability and
simpler implementation by decoupling the primary functions of an
application-aware MC into a simpler multi-stage MC, and 2) performance and
fairness improvement by reducing the interference \changesII{caused by very} bandwidth-intensive
applications. \titleShortSMS provides these benefits by introducing
a three-stage design. The first stage is the \changesII{per-core} \emph{\batch formation} stage, \changesI{which}
groups requests from the same application that access the same row to improve
\mbox{row buffer} locality. The second stage is the \emph{\batch scheduler}, \changesI{which} 
schedules batches of requests \changesII{from} across different applications. The last
stage is the \emph{DRAM command scheduler}, \changesII{which sends requests to DRAM} while
satisfying all DRAM constraints.

\begin{figure}[h]
\centering
\includegraphics[width=\columnwidth]{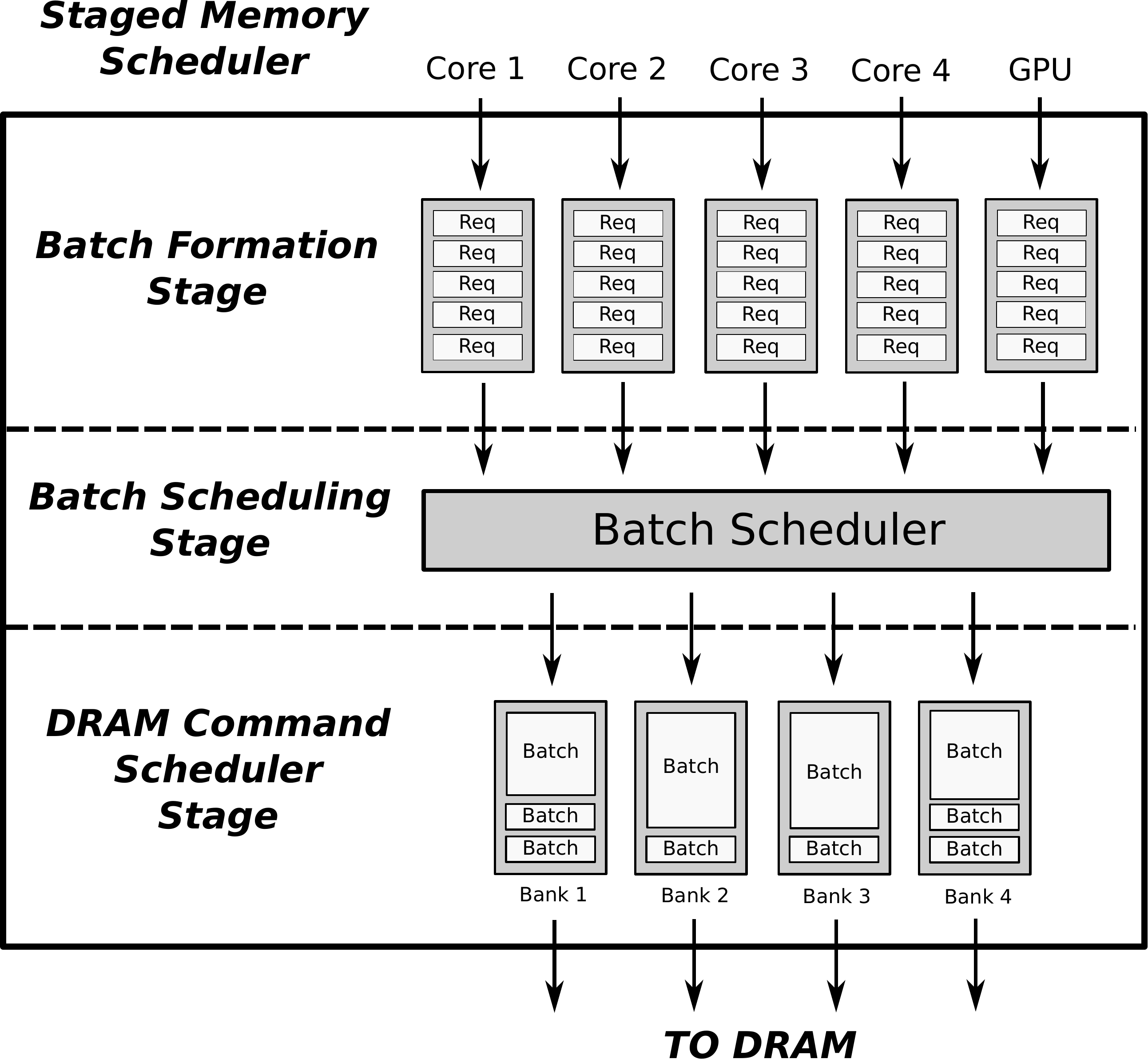}
\caption{The organization of \titleShortSMS. Adapted from~\cite{sms}.}
\label{fig:overall}
\end{figure}

\sloppypar{The staged organization of \titleShortSMS lends directly to a \mbox{low-complexity}
hardware implementation. Figure~\ref{fig:overall} illustrates the overall
hardware organization of the \titleShortSMS. \changesI{We briefly discuss each stage below. 
Section~4 of our ISCA 2012 paper~\cite{sms} includes a detailed description of each stage.}}

\label{sec:algoSMS}
{\bf Stage 1 - \Batch Formation.}
The goal of this stage is to combine individual memory requests from each source
into batches of \changesII{requests that are to the same row buffer entry}.
It consists of several simple FIFO structures, one
per {\em source} (i.e., a CPU core or the GPU). Each request from a given
source is initially inserted into its respective FIFO upon arrival at the
MC. A {\em\batch} is simply one or more memory requests from
the same source that access the same DRAM row. That is, all requests within a
\batch, except perhaps for the first one, would be row buffer hits if scheduled
consecutively. A \batch is deemed complete or {\em ready} when an incoming request
accesses a different row, when the oldest request in the \batch has exceeded a
threshold age, or when the FIFO is full. Only ready batches are considered \changesII{for future scheduling}
by the second stage of \titleShortSMS.

{\bf Stage 2 - \Batch Scheduler.}
\ignore{
The \batch formation stage has combined individual memory requests into \batches of
row buffer hitting requests.}
The \batch scheduler deals directly with \batches, and therefore \changesI{does not need to worry}
about \changesII{optimizing} for \mbox{row buffer} locality. Instead, the
\batch scheduler focuses on higher-level policies regarding inter-application
interference and fairness. The goal of this stage is to prioritize \batches from applications
that are latency critical, while making sure that bandwidth-intensive
applications (e.g., \changesI{those running on} the GPU) still make \changesII{good} progress.

The \batch scheduler considers every source FIFO (from stage 1) that
contains a ready batch. It picks one ready batch based on either a shortest job
first (SJF) or a round-robin policy.  Using the SJF policy, the \batch
scheduler chooses the oldest ready batch from the source with the fewest total
in-flight memory requests across all three stages of \titleShortSMS. 
SJF prioritization reduces average request service latency, and it tends to favor
latency-sensitive applications, which tend to have fewer total requests~\cite{tcm,atlas,parbs,cjlee-micro09}. 
Using the round-robin policy, the \batch scheduler simply picks the next ready batch
in a \mbox{round-robin} manner across the source FIFOs. This ensures that
\changesII{memory-intensive} applications receive adequate service. The \batch
scheduler uses the SJF policy with probability $p$ and the round-robin policy
with probability $1-p$. The value of $p$ determines whether the CPU or the GPU
receives higher priority. When $p$ is high, the SJF policy is applied more
often and applications with fewer outstanding requests are prioritized. Hence,
the batches of the likely less memory-intensive CPU applications are
prioritized over the batches of the GPU application. On the other hand, when
$p$ is low, request batches are scheduled in a round-robin fashion more often.
Hence, the memory-intensive GPU application's \changesII{naturally-large} request batches are likely
scheduled more frequently, and the GPU is \changesII{thus} prioritized over the CPU.

%
After picking a \batch, the \batch scheduler enters a drain state where it
forwards the requests from the selected \batch to the final stage of the
\titleShortSMS. The \batch scheduler dequeues one request per cycle until
all requests from the \batch have been removed from the selected FIFO.

\begin{figure*}
\centering
\includegraphics[width=\textwidth]{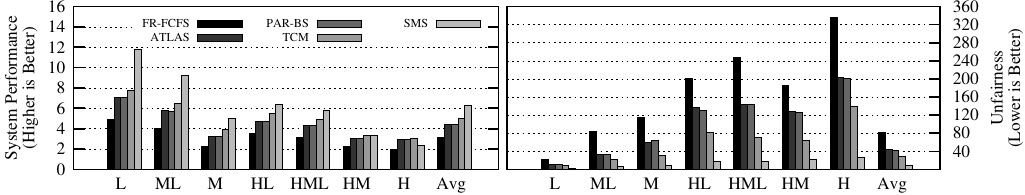}
\caption{System performance, and fairness for 7 categories of workloads (total
of 105 workloads). Reproduced from~\cite{sms}.}
\label{fig:mainres}
\end{figure*}

\begin{figure*}
\centering
\includegraphics[width=\textwidth]{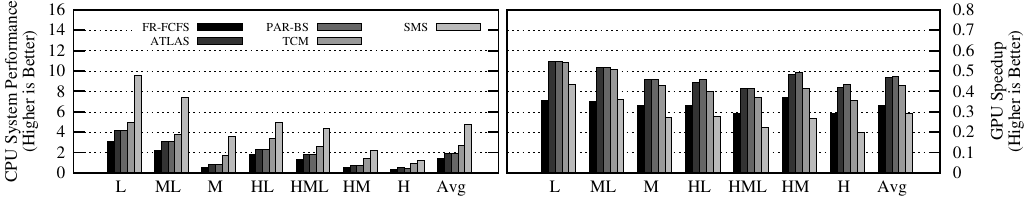}
\caption{CPUs and GPU Speedup for 7 categories of workloads (total
of 105 workloads). Reproduced from~\cite{sms}.}
\label{fig:spdUp}
\end{figure*}

{\bf Stage 3 - DRAM Command Scheduler (DCS).} DCS consists of one
FIFO queue per DRAM bank. The drain state of the \batch scheduler places the
memory requests directly into these FIFOs. Note that because \batches are moved
into DCS FIFOs one \batch at a time, \mbox{row buffer} locality within a
\batch is preserved within a DCS FIFO. At this point, higher-level policy
decisions have already been made by the \batch scheduler. Therefore, the DCS
simply issues low-level DRAM commands, ensuring DRAM protocol compliance.

In any given cycle, DCS considers only the requests at the {\em head} of
each of the per-bank FIFOs. For each request, DCS determines whether that
request can issue a command based on the request's current row buffer state
(e.g., is the row buffer already open with the requested row?) and the current
DRAM state (e.g., time elapsed since a row was opened in a bank, and data bus
availability). If more than one request is eligible to issue a command in any
given cycle, the DCS arbitrates between DRAM banks in a round-robin fashion.


\section{Qualitative Comparison with\\ Previous Scheduling Algorithms}
\label{sec:sms-qual-eval}
In this section, we compare \titleShortSMS qualitatively to previously proposed
scheduling policies and analyze the basic differences between \titleShortSMS and
these policies. The fundamental difference between \titleShortSMS and previously-proposed 
memory scheduling policies for \changesII{CPU-only} scenarios is that the latter
are designed around a single, centralized request buffer which has poor
scalability and complex scheduling logic, while \titleShortSMS is built around a
decentralized, scalable framework.

\paragraphbe{First-Ready FCFS (FR-FCFS).}
FR-FCFS~\cite{fr-fcfs,frfcfs-patent} is a commonly used scheduling
policy in commodity DRAM systems. An FR-FCFS scheduler prioritizes requests
that result in row buffer hits over row buffer misses and otherwise prioritizes
older requests. Since FR-FCFS unfairly prioritizes applications with high
row buffer locality to maximize DRAM throughput, prior
works~\cite{atlas,parbs,tcm,stfm,memattack,bliss,bliss-tpds,a2c,mcp,fst} have observed that it has low system
performance and high unfairness.

\paragraphbe{Parallelism-Aware Batch Scheduling (PAR-BS).}
PAR-BS~\cite{parbs,parbs-tp} aims to improve \changesII{both}
fairness and system performance. In order to prevent unfairness, it forms
batches of outstanding memory requests and prioritizes the oldest batch, to
avoid request starvation. To improve system throughput, it prioritizes
applications with smaller number of outstanding memory requests within a batch.
However, PAR-BS has two major shortcomings. First, batching could cause older
GPU requests and requests of other memory-intensive CPU applications to be
prioritized over latency-sensitive CPU applications. Second, as previous
work~\cite{atlas} has also observed, PAR-BS does not take into account an
application's long term memory-intensity characteristics when it assigns
application priorities within a batch. This could cause memory-intensive
applications' requests to be prioritized over latency-sensitive applications'
requests within a batch, \changesII{due to the application-agnostic nature of batching}.

\paragraphbe{Adaptive Per-Thread Least-Attained-Serviced Memory Scheduling (ATLAS).}
ATLAS~\cite{atlas} aims to improve system performance by prioritizing
requests of applications with lower attained memory service. This improves the
performance of low memory-intensity applications as they tend to have low
attained service. However, ATLAS has the disadvantage of not preserving
fairness. Previous works~\cite{atlas,tcm} have shown that simply prioritizing
applications \changesII{based on attained service} leads to significant slowdown of memory-intensive
applications.

\paragraphbe{Thread Cluster Memory Scheduling (TCM).}
TCM~\cite{tcm} is a
state-of-the-art application-aware \changesII{cluster} memory scheduler providing both \changesII{high system
throughput and high fairness. It groups an application into either a latency-sensitive or
a bandwidth-sensitive cluster based on the application memory intensity.} In order to
achieve high system throughput and low unfairness, TCM employs a different
prioritization policy for each cluster. To improve system throughput, a
fraction of total memory bandwidth is dedicated to the latency-sensitive cluster
and applications within the cluster are then ranked based on memory intensity
with \changesII{the} least memory-intensive application receiving the highest priority. On the
other hand, TCM minimizes unfairness by periodically shuffling applications
within \changesII{the} bandwidth-sensitive cluster to avoid starvation. This approach
provides both high system performance and fairness in CPU-only systems. In an
integrated CPU-GPU system, \changesII{the} GPU generates a significantly larger \changesII{number} of
memory requests compared to \changesII{the} CPUs and fills up the centralized request buffer. 
As a result, the memory controller lacks the visibility \changesII{into} CPU memory requests
to accurately determine each application's memory access \changesII{characteristics}. Without \changesII{such}
visibility, TCM makes incorrect and non-robust clustering decisions, which 
classify some applications with high memory intensity into the
latency-sensitive cluster \changesII{and vice versa}. \changesII{Such} misclassified applications cause
interference not only to low memory intensity applications, but also to each
other. Therefore, TCM \changesII{cannot always provide high} 
system performance and \changesII{high} fairness in an integrated CPU-GPU 
system. Increasing the request buffer size is a 
\changesII{practical} way to gain more visibility into CPU applications' memory
access \changesII{characteristics}. However, this approach is not scalable as we show in our
evaluations~\cite{sms}. In contrast, \titleShortSMS provides much
better system performance and fairness than TCM with the same number of
request buffer entries and lower hardware cost, \changesIII{as we show in Section~\ref{sec:results}}.

\section{Evaluation Methodology}
\label{sec:meth}

We use an in-house cycle-accurate simulator to perform our evaluations. For our
performance evaluations, we model a system with sixteen x86 CPU cores and a
GPU. For the CPUs, we model three-wide out-of-order processors with a cache
hierarchy including per-core L1 caches and a shared, distributed L2 cache.  The
GPU does not share the CPU caches.  In order to prevent the GPU from taking the
majority of request buffer entries, we reserve half of the request buffer
entries for the CPUs. To model the memory bandwidth of the GPU accurately, we
perform coalescing on GPU memory requests before they are sent to the memory
controller~\cite{lindholm}.

We evaluate our system with a set of 105 multiprogrammed workloads 
simulated for 500~million cycles.  \changesII{Each} workload consists of
sixteen SPEC CPU2006 benchmarks and one GPU application selected from a mix of
video games and graphics performance benchmarks. We classify CPU benchmarks
into three categories (Low, Medium, and High) based on their memory
intensities, measured as last-level cache misses per thousand instructions
(MPKI). Based on these three categories, we randomly choose sixteen CPU
benchmarks from these three categories and one randomly selected GPU benchmark
to form workloads consisting of seven intensity mixes: L (All low), ML
(Low/Medium), M (All medium), HL (High/Low), HML (High/Medium/Low), HM
(High/Medium) and H(All high). For each CPU benchmark, we use
Pin~\cite{reddi2004pin,pin} with PinPoints~\cite{pinpoint} to select the
representative phase.  For the GPU applications, we use an industrial GPU
simulator to collect memory requests with detailed timing information.  These
requests are collected after having first been filtered through the GPU's
internal cache hierarchy, therefore we do not further model any caches for the
GPU in our final hybrid CPU-GPU simulation framework.  \changesII{More detail
on our experimental methodology is in Section~5 of our ISCA 2012
paper~\cite{sms}.}

\begin{figure*}
\centering
\includegraphics[width=2\columnwidth]{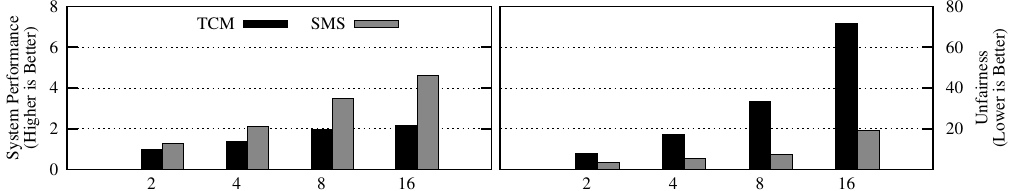}
\caption{\titleShortSMS vs. TCM on a 16 CPU/1 GPU, 4 memory controller system with varying the number of cores.}
\label{fig:coreSweep}
\end{figure*}

\begin{figure*}
\centering
\includegraphics[width=2\columnwidth]{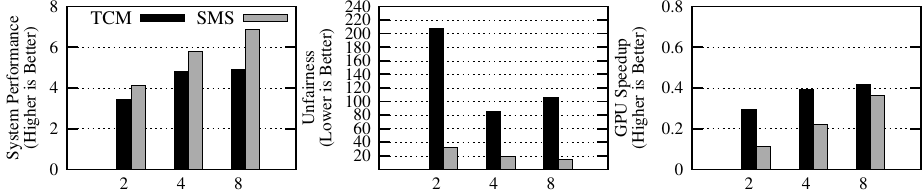}
\caption{\titleShortSMS vs. TCM on a 16 CPU/1 GPU system with varying the number of channels.}
\label{fig:chanSweep}
\end{figure*}

\section{Experimental Results}
\label{sec:results}

We present the performance of five memory scheduler configurations:
FR-FCFS~\cite{fr-fcfs,frfcfs-patent}, ATLAS~\cite{atlas}, PAR-BS~\cite{parbs}, TCM~\cite{tcm}, and SMS~\cite{sms} on the 16-CPU/1-GPU four-memory-controller system. 
All memory schedulers use 300 request
buffer entries per memory controller. This size was chosen based on \changesII{empirical} results, which showed that performance does not appreciably
increase for larger request buffer sizes. Results are presented in the
workload categories, with
workload memory intensities increasing from left to right.

Figure~\ref{fig:mainres} shows the system performance (measured as weighted
speedup~\cite{harmonic_speedup,ws-metric2}) and fairness (measured as maximum slowdown~\cite{Reetu-MICRO2009,atlas,tcm,vandierendonck,bliss,bliss-tpds}) 
of the previously proposed algorithms and \titleShortSMS.
Compared to TCM, which is the previous state-of-the-art algorithm for
both system performance and fairness, \titleShortSMS provides 41.2\% system
performance improvement and 4.8$\times$ fairness improvement. Therefore, we
conclude that \titleShortSMS provides better system performance and fairness
than all previously proposed scheduling policies, while incurring much lower
hardware cost and simpler scheduling logic, as we show in Section~\ref{sec:hardware}.

We study the performance of the CPU system and the GPU system separately and
provide two major observations in Figure~\ref{fig:spdUp}. First, \titleShortSMS gains
1.76$\times$ improvement in CPU system performance over TCM. Second,
\titleShortSMS achieves this 1.76$\times$ CPU performance improvement while
delivering similar GPU performance as the FR-FCFS baseline. 
The results show that TCM (and the other algorithms) end up
allocating far more bandwidth to the GPU, at significant performance and
fairness cost to the CPU applications.  \titleShortSMS appropriately
deprioritizes the memory bandwidth intensive GPU application in order to enable
higher CPU performance and overall system performance, while preserving
fairness. Previously proposed scheduling algorithms, on the other hand, allow
the GPU to hog memory bandwidth and therefore significantly degrade system performance
and fairness.

\changesII{We provide a more detailed analysis in Sections~6.1 and 6.2 of our 
ISCA 2012 paper~\cite{sms}.}

\subsection{Scalability with Cores and\\Memory Controllers}

Figure~\ref{fig:coreSweep} compares the performance and fairness of \titleShortSMS
against TCM (averaged over 75 workloads)\footnote{We use 75 randomly selected workloads
per core count.  We could not use the same workloads/categorizations as specified in Section~\ref{sec:meth} because
those were for 16-core systems, whereas we are now varying the number of cores.} with the same number of
request buffers, as the number of cores is varied. We make the following
observations: First, \titleShortSMS continues to provide better system performance
and fairness than TCM. Second, the system performance gains and
fairness gains increase significantly as the number of cores and
hence, memory pressure is increased. \titleShortSMS's performance and
fairness benefits are likely to become more significant as core counts in future
technology nodes increase.

Figure~\ref{fig:chanSweep} shows the system performance and fairness of
\titleShortSMS compared against TCM as the number of memory channels is varied.
For this, and all subsequent results, we perform our evaluations on 60 workloads
from categories that contain high memory-intensity applications (HL, HML, HM and H workload
categories). We observe that
\titleShortSMS scales better as the number of memory channels increases. As the
performance gain of TCM diminishes when the number of memory channels increases
from 4 to 8 channels, \titleShortSMS continues to provide performance improvement
for both CPU and GPU. \changesII{We provide a detailed scalability analysis in Section~6.3 
of our ISCA 2012 paper~\cite{sms}.}

\subsection{Power and Area}
\label{sec:hardware}

\changesII{We present the power and area of FR-FCFS and SMS. We find that SMS consumes
66.7\% less leakage power than FR-FCFS, which is the simplest of all of the prior
memory schedulers that we evaluate. In terms of die area, SMS requires 46.3\% less area than
FR-FCFS. The majority of the power and area savings of SMS over FR-FCFS come
from the decentralized request buffer queues and simpler scheduling logic in
SMS.  In comparison, FR-FCFS requires centralized request buffer queues, content-addressable
memory (CAMs), and complex scheduling logic. Because ATLAS and TCM
require more complex ranking and scheduling logic than FR-FCFS, we expect that SMS
also provides power and area reductions over ATLAS and TCM.}

\changesII{We provide the following additional results in our ISCA 2012 paper~\cite{sms}:}

\begin{itemize}
\item Combined performance of CPU-GPU heterogeneous systems for different \titleShort configurations with different
      Shortest Job First (SJF) probability.
\item Sensitivity analysis to \titleShort's configuration parameters.
\item Performance of \titleShort in CPU-only systems.
\end{itemize}

\section{Related Work}

\changesII{To our knowledge, \changesIII{our ISCA 2012 paper is the first} to 
provide a fundamentally
new memory controller design for heterogeneous CPU-GPU systems in order to
reduce interference at the shared off-chip main memory.}
\changesIII{There are several prior works that} reduce interference at the shared off-chip
main memory \changesIII{in other systems}. 
We provide a brief discussion of these works.

\subsection{Memory Partitioning Techniques}

Instead of mitigating the interference problem between applications by
scheduling requests at the memory controller, Awasthi et al.~\cite{rajeev-pact10} propose a
mechanism that spreads data in the same working set across memory channels in
order to increase memory level parallelism. Memory channel partitioning (MCP)~\cite{mcp} maps applications to different
memory channels based on their memory intensities and row buffer locality, to
reduce inter-application interference. Mao et al.~\cite{mao-temp} propose to
partition GPU channels and allow only a subset of threads to access each memory
channel. In addition to channel partitioning, several works~\cite{xie-hpca14,ikeda-icpads13,liu-pact12}
also propose to partition DRAM
banks to improve performance. These
partitioning techniques are orthogonal to our proposals, and can be combined with \titleShort to
improve the performance of heterogeneous CPU-GPU systems.

\subsection{Memory Scheduling Techniques} 

\paragraphbe{Memory Scheduling on CPUs.}
Numerous prior works propose memory scheduling algorithms for CPUs that
improve system performance.
The first-ready, first-come-first-serve (FR-FCFS) scheduler~\cite{fr-fcfs, frfcfs-patent}
prioritizes requests that hit in the row buffer over requests that miss in the row buffer,
with the aim of reducing the number of times rows must be activated (as row activation
incurs a high latency).
Several memory schedulers improve performance beyond FR-FCFS by
identifying critical threads in multithreaded applications~\cite{pam},
using reinforcement learning to identify long-term memory behavior~\cite{ipek-isca08, morse-hpca12},
prioritizing memory requests based on the criticality (i.e., latency sensitivity) of
each memory request~\cite{ghose2013, xiong-taco16, liu-ipccc16},
distinguishing prefetch requests from demand requests~\cite{pa-micro08, cjlee-micro09},
or improving the scheduling of memory writeback requests~\cite{vwq-isca10, lee2010dram, dbi}.
While all of these schedulers increase DRAM performance and/or throughput,
many of them introduce fairness problems by under-servicing applications
that only infrequently issue memory requests.
To remedy fairness problems, several application-aware memory scheduling
algorithms~\cite{mutlu-podc08,atlas,tcm,stfm,parbs,bliss,bliss-tpds,mise} use information
on the memory intensity of each application to balance both performance and fairness. 
Unlike SMS, none of these schedulers consider the different needs of CPU memory requests and
GPU memory requests in a heterogeneous system.

\paragraphbe{Memory Scheduling on GPUs.}
Since GPU applications are bandwidth intensive, often with streaming access
patterns, a policy that maximizes the number of row buffer hits is effective
for GPUs to maximize overall throughput. As a result, FR-FCFS with a large
request buffer tends to perform well for GPUs~\cite{gpgpu-sim}. In view of this,
prior work~\cite{complexity} proposes mechanisms to reduce the complexity of
FR-FCFS scheduling for GPUs. Ausavarungnirun et al.~\cite{medic} propose MeDiC, which
is a cache and memory management scheme to improve the performance of GPGPU applications. 
Jeong et al.~\cite{jeong2012qos} propose a QoS-aware
memory scheduler that guarantees the performance of GPU applications by
prioritizing memory requests from graphics applications over 
those from CPU applications until the system can
guarantee that a frame can be rendered within a given deadline, after which it prioritizes
requests from CPU applications. Jog et
al.~\cite{adwait-critical-memsched} propose CLAM, a memory scheduler that
identifies critical memory requests and prioritizes them in the main memory.
Ausavarungnirun et al.~\cite{mask} propose a scheduling algorithm
that identifies and prioritizes TLB-related memory requests in GPU-based systems,
to reduce the overhead of memory virtualization.
Unlike \titleShort, none of these works holistically optimize the performance \emph{and}
fairness of requests when a memory controller is shared by a CPU and a GPU.

\paragraphbe{Memory Scheduling on Emerging Systems.}
Recent proposals investigate memory
scheduling on emerging platforms. Usui et al.~\cite{usui-squash, usui-dash}
propose
accelerator-aware memory controller designs that improve the performance of
systems that contain both CPUs and hardware accelerators. 
Zhao et al.~\cite{jishen-firm} decouple the design of a memory controller for persistent memory into multiple
stages.
These works build upon principles for heterogeneous system
memory scheduling that were first proposed in \titleShort.

\subsection{Other Related Works}

\paragraphbe{DRAM Designs.}
Aside from memory scheduling and memory partitioning techniques, previous works
propose new DRAM designs that are capable of reducing memory latency in
conventional
DRAM~\cite{chang-sigmetric16,lisa,dsarp,al-dram,tl-dram,ava-dram,donghyuk-stack,salp,lee-pact2015,micron-rldram3,sato-vlsic1998,hart-compcon1994,hidaka-ieeemicro90,hsu-isca1993,kedem-1997,son-isca2013,luo-dsn2014,chatterjee-micro2012,phadke-date2011,shin-hpca2014,chandrasekar-date2014,o-isca2014,zheng-micro2008,ware-iccd2006,ahn-cal2009,ahn-taco2012}
and non-volatile
memory~\cite{meza-weed2013,ku-ispass2013,meza-cal2012,yoon-iccd2012,qureshi-isca2009,qureshi-micro2009,lee-isca2009,lee-ieeemicro2010,lee-cacm2010}.
Previous works on bulk data transfer~\cite{gschwind-cf2006, gummaraju-pact2007,
kahle-ibmjrd2005,carter-hpca1999,zhang-ieee2001,seo-patent,intelioat,zhao-iccd2005,jiang-pact2009,seshadri2013rowclone,lu-micro2015,lisa}
and in-memory
computation~\cite{ahn-isca2015,ahn-isca2015-2,7056040,7429299,guo-wondp14,592312,
seshadri-cal2015,mai-isca2000,draper-ics2002,
seshadri-micro2015,ambit,hsieh-iccd2016,tom-isca16,
amirali-cal2016, stone-1970, fraguela-2003,375174,808425,
4115697,694774,sura-2015,zhang-2014,akin-isca2015,
babarinsa-2015,7446059,6844483,pattnaik-pact2016, kim.bmc18,
boroumand.asplos18} can be used improve DRAM
bandwidth.  Techniques to reduce the overhead of DRAM refresh~\cite{raidr,
venkatesan-hpca2006,bhati-isca2015, lin-iccd2012, agrawal-memsys2016,
nair-isca2013,
kim-asic2001,baek-tc2014,agrawal-hpca2014,ohsawa-islped1998,qureshi-dsn2015} can
be applied to improve the performance of GPU-based systems.
Data compression techniques~\cite{bdi-pact12,lcp-micro13,toggle-hpca16,compress-reuse-hpca15,caba} 
can also be used on the main memory to increase the
effective available DRAM bandwidth.
All of these techniques can mitigate the performance impact of memory
interference and improve the performance of GPU-based systems. They are
orthogonal to, and can be combined with, \titleShort to further improve the performance
of heterogeneous CPU-GPU systems.

Previous works on data prefetching~\cite{pa-micro08,seshadri-taco2015,cjlee-micro09,nesbit-pact2004,srinath-hpca2007,lai-isca2001,alameldeen-hpca2007,baer-1995,cao-sigmetrics1995,dahlgren-1995,ebrahimi-micro09,jouppi-isca90,hur-micro2006,joseph-isca1997,cooksey-asplos2002,ebrahimi-hpca,ebrahimi-isca2011,mutlu-isca2005,
mutlu-hpca2003, mutlu-micro2005,
hashemi-isca2016,lee-tc2011,hashemi-micro2016} can also be used to mitigate high DRAM
latency. However, these techniques generally
increase DRAM bandwidth utilization, which can lead to lower GPU performance.

\paragraphbe{Other Ways to Improve Performance on Systems with GPUs.} 
Other works have
proposed various methods of decreasing memory
divergence.
These methods range from thread throttling~\cite{ccws,nmnl-pact13,cpugpu-micro,kuo-throttling14} to warp
scheduling~\cite{ccws,largewarps,warpsub,zheng-cal,caws-pact14,tor-micro13}.
While these methods share our goal of reducing memory divergence, none of them
exploit \emph{inter-warp} heterogeneity and, as a result, are orthogonal or
complementary to our proposal.  Our work also makes new observations about 
memory divergence that are not covered by these works.

\section{Significance and Long-Term Impact}
\label{sec:sms-impact}


%
%


\titleShortSMS exposes the need to redesign components of the memory subsystem
to better serve integrated CPU-GPU systems.  Systems-on-chip (SoCs) that
integrate CPUs and GPUs on the same die are growing rapidly in popularity
(e.g., \cite{arm-mali,amdzen,tegra,tegrax1}), due to
their high energy efficiency and lower costs compared to discrete CPUs and GPUs.  
As a result, SoCs are commonly used in mobile devices
such as smartphones, tablets, and laptops, and are being used in many servers
and data centers.
We expect that as more powerful CPUs and GPUs are integrated in SoCs, and
as the workloads running on the CPUs/GPUs become more memory-intensive,
\titleShortSMS will become even more essential to alleviate the shared memory
subsystem bottleneck.

The observations and mechanisms in our ISCA 2012 paper~\cite{sms}
expose several future research problems.  We briefly discuss two
future research areas below.


\paragraphbe{Interference Management in Emerging Heterogeneous Systems.} 
Our ISCA 2012 paper~\cite{sms} considers heterogeneous systems where
a CPU executes various general-purpose applications while the GPU executes graphics
workloads.  Modern heterogeneous systems contain an increasingly diverse set
of workloads.  For example, programmers can use the GPU in an integrated
CPU-GPU system to execute general-purpose applications (known as GPGPU
applications).  GPGPU applications can have significantly different access patterns
from graphics applications, requiring different memory scheduling policies
(e.g., \cite{adwait-critical-memsched, medic, mask}).  Future work can
adapt the mechanisms of \titleShortSMS to optimize the performance of
GPGPU applications.

Many heterogeneous systems are being deployed in mobile or embedded environments,
and must ensure that memory requests from some or all of the components of the heterogeneous system
meet \emph{real-time deadlines}~\cite{kim-rtas2014, kim2016, usui-squash, usui-dash}.
Traditionally, applications with real-time deadlines are executed using embedded cores
or fixed-function accelerators, which are often integrated into modern SoCs.
We believe that the observations and mechanisms in our ISCA 2012 paper~\cite{sms}
can be used and extended to ensure that these deadlines are met.
Recent works~\cite{usui-squash, usui-dash} have shown that the principles
of \titleShortSMS can be extended to provide deadline-aware memory scheduling for accelerators
within heterogeneous systems.


Even though the mechanisms proposed in our ISCA 2012 paper~\cite{sms} aim to minimize the
slowdown caused by interference, they do not provide actual
performance guarantees. However, we believe it is possible to 
combine principles from \titleShortSMS with
prediction mechanisms for memory access latency (e.g., \cite{kim-rtas2014, kim2016, mise,asm-micro15}
to provide \emph{hard} performance guarantees for real-time applications, while still
ensuring fairness
for all applications executing on the heterogeneous system.

\paragraphbe{Memory Scheduling for Concurrent GPGPU Applications.}
While \titleShortSMS allows CPU applications and graphics applications to share
DRAM more efficiently, we assume that there is only a single GPU application
running at any given point in time. \changesI{Recent works~\cite{mosaic,mafia} propose
methods to efficiently share the same GPU across multiple concurrently-executing
GPGPU applications. We believe that the
techniques and observations provided in our ISCA 2012 paper~\cite{sms} 
can be applied to reduce the memory interference induced by additional GPGPU applications.
Furthermore, as concurrent GPGPU application execution becomes more widespread,
the concepts of \titleShortSMS can be extended to provide prioritization and fairness
across multiple GPGPU applications.}

\changesIII{
Our analysis of memory interference in heterogeneous systems, and
our new \titleLongSMS, have inspired a number of subsequent works.
These works include significant research on
predictable and/or deadline-aware memory
scheduling~\cite{kim-rtas2014, kim2016, usui-squash, usui-dash,
jishen-firm,bliss,bliss-tpds,mise}, and on other memory scheduling 
algorithms for heterogeneous
systems~\cite{usui-squash,usui-dash,wang-pact14,pattnaik-pact2016}}.

\section{Conclusion}

While many advancements in memory scheduling policies have been made to deal
with multi-core processors, the integration of GPUs on the same chip as the
CPUs has created new system design challenges.  Our ISCA 2012 paper~\cite{sms} demonstrates
how the inclusion of GPU memory traffic can cause severe difficulties for
existing memory controller designs in terms of performance and especially fairness. 
We propose a new approach, \titleLongSMS, which delivers superior
performance and fairness for integrated CPU-GPU systems compared to state-of-the-art memory schedulers,
while providing a design that is significantly simpler to implement (thus
improving the scalability of the memory controller). The key
insight behind simplifying the implementation of \titleShortSMS is that the primary functions of
sophisticated memory controller algorithms can be decoupled.
As a result, \titleShortSMS proposes a multi-stage memory controller
architecture.
We show that \titleShortSMS significantly improves the performance and fairness
in integrated CPU-GPU systems.
We hope and expect that our observations and mechanisms can inspire future work
in memory system design for existing and emerging heterogeneous systems.


\section*{Acknowledgments}

We thank Saugata Ghose for his dedicated effort in the preparation
of this article.
We thank Stephen Somogyi and Fritz Kruger at AMD for their
assistance with the modeling of the GPU applications. We also
thank Antonio Gonzalez, anonymous reviewers and members
of the SAFARI group at CMU for their feedback. We acknowledge
the generous support of AMD, Intel, Oracle, and
Samsung. This research was also partially supported by grants
from the NSF (CAREER Award CCF-0953246 and CCF-1147397),
GSRC, and Intel ARO Memory Hierarchy Program.

{
\bibliographystyle{IEEEtranS}
\bibliography{references}
}

\end{document}